\documentclass[prc,12pt,tightenlines,amsfonts,nofootinbib]{revtex4}

\usepackage{amssymb}
\usepackage{bm}

\begin{document}
%
%
%
%
%
%
%
%
%
%
%
\let\dsize=\displaystyle
\let\tsize=\textstyle
\let\ssize=\scriptstyle
\let\sssize\scriptscriptstyle
%
%
%
%
%
\def\spacesymbol{\leavevmode\hbox{\tt\char'040}}
\def\tildesymbol{\mathchar"0218 }
\def\coeff#1#2{{\textstyle{#1\over #2}}}
\def\tover#1#2{{\textstyle {#1\over #2}}}
\def\shalf{\tover12}
\def\e{\,{\rm e}}
\def\Tr{\mathop{\rm Tr}\nolimits}
\def\tr{\mathop{\rm tr}\nolimits}
\def\lsim{\lower0.6ex\vbox{\hbox{$\ \buildrel{\textstyle <}
         \over{\sim}\ $}}}
\def\gsim{\lower0.6ex\vbox{\hbox{$\ \buildrel{\textstyle >}
         \over{\sim}\ $}}}
%
\def\Order#1{{\cal O}(#1)}
\def\pairof#1{#1^+ #1^-}
\def\ee{\pairof{e}}
\def\CM{{\rm CM}}
\def\bar#1{\overline{#1}}

\def\dAlembertian{\frame{\phantom{\rule{8pt}{8pt}}}}
\let\bo=\dAlembertian
\def\dell{\bigtriangledown}   

\def\th{$^{\rm th}$}
\def\st{$^{\rm st}$}
\def\nd{$^{\rm nd}$}
\def\rd{$^{\rm rd}$}

\def\unittie{\nobreak\,}
\def\fm{\unittie\hbox{fm}}
\def\fminverse{\unittie\hbox{fm}^{-1}}          \let\infm=\fminverse
\def\eV{\unittie\hbox{eV}}    \let\ev=\eV
\def\MeV{\unittie\hbox{MeV}}    \let\Mev=\MeV  \let\mev=\MeV
\def\GeV{\unittie\hbox{GeV}}    \let\Gev=\GeV  \let\gev=\GeV
\def\MeVfm{\hbox{MeV--fm}}  \let\mevfm=\MeVfm

%
%
%

\def\undervec#1{\setbox0\hbox{$\tildesymbol$}\setbox1\hbox{$#1$}#1%
                \dimen0=\wd0\advance\dimen0by\wd1\divide\dimen0by2
                 \kern-\dimen0\lower1.3ex\hbox{$\tildesymbol$}}
\def\sundervec#1{\setbox0\hbox{$\ssize\tildesymbol$}
                 \setbox1\hbox{$\ssize #1$}#1%
                \dimen0=\wd0\advance\dimen0by\wd1\divide\dimen0by2
                 \kern-\dimen0\lower.85ex\hbox{$\ssize\tildesymbol$}}
\def\ssundervec#1{\setbox0\hbox{$\sssize\tildesymbol$}
                 \setbox1\hbox{$\sssize #1$}#1%
                \dimen0=\wd0\advance\dimen0by\wd1\divide\dimen0by2
                 \kern-\dimen0\lower.6ex\hbox{$\sssize\tildesymbol$}}

\def\uvec#1{{\mathchoice{\undervec#1}%
     {\undervec#1}{\sundervec#1}{\ssundervec#1}}}

\def\longvec{\overrightarrow}

\def\boldvector#1{\ifcat#1A {{\bf #1}} \else \mbf{#1} \fi}

\def\mbf#1{{
\mathchoice
{\hbox{\boldmath$\displaystyle{#1}$}}
{\hbox{\boldmath$\textstyle{#1}$}}
{\hbox{\boldmath$\scriptstyle{#1}$}}
{\hbox{\boldmath$\scriptscriptstyle{#1}$}}
}}

\def\vectorsunder{\def\vector{\uvec}}
\def\vectorsbold{\def\vector{\boldvector}}      
\def\vectorsarrow{\def\vector{\mathaccent"017E }}  
\vectorsbold

%
%
%
\def\Tc{T_c}
\def\kfermi{k_{\sssize {\rm F}}}    
\def\epsfermi{\epsilon_{\sssize {\rm F}}}    
     \let\efermi=\epsfermi
\def\Mstar{M^{\ssize\ast}}       
\def\Mstarsq{M^{\ssize\ast}{}^2}
\def\Estar{E^{\ssize\ast}}
\def\EFstar{E^{\ssize\ast}_{\sssize {\rm F}}}
     \let\Efstar=\EFstar
\def\kstar{k^{\ssize\ast}}
\def\rhosmall#1{\rho_{\sssize {#1}}}
\def\rhoB{\rhosmall{\rm B}}
\def\rhozero{\rhosmall{0}}
\def\rhothree{\rhosmall{3}}
\def\rhos{\rho_{{\rm s}}}
\def\rhov{\rho_{{\rm v}}}
\def\Ylzero{Y_{l0}}     \let\Ylzer=\Ylzero
\def\Ylm{Y_{lm}}       \let\ylm=\Ylm
\def\Ghartree{G_{{\ssize \rm H}}}   \let\GHartree=\Ghartree
\def\Msun{M_{\odot}}
%
%
%
%
%
\def\mn{{\mu\nu}}         \let\munu=\mn
\def\gmunu{g^\mn}         \def\etamunu{\eta^\mn}
\def\gmunul{g_\mn}        \def\etamunul{\eta^\mn}
\def\sigmamunu{\sigma^\mn}
\def\sigmamunul{\sigma_\mn}
\def\sigmai{\sigma^i}
\def\sigmaj{\sigma^j}
\def\sigmaij{\sigma^{ij}}
\def\sigmavec{\vector\sigma}
\def\tauthree{\tau_{{\sssize 3}}}
\def\tauvec{\vector\tau}
\def\gammamu{\gamma^{\mu}}       \let\gammu=\gammamu
\def\gammamul{\gamma_{\mu}}      \let\gammul=\gammamul
\def\gammafive{\gamma^5}         \let\gamfiv=\gammafive
\def\gammafivel{\gamma_5}        \let\gamfivl=\gammafivel
\def\gammazero{\gamma^{0}}       \let\gamzer=\gammazero
\def\gammazerol{\gamma_{0}}      \let\gamzerl=\gammazerol
\def\gammavec{\vector\gamma}        \let\gamvec=\gammavec
\def\gammavecl{\vector\gamma}       \let\gamvecl=\gammavecl
\def\alphavec{\vector\alpha}         \let\alphvec=\alphavec
\def\slash#1{\rlap/{#1}}
\def\kslash{\slash{\mkern-1mu k}}    
%
\def\slashi#1{\rlap{\sl/}#1}
%
\def\slashii#1{\setbox0=\hbox{$#1$}             
   \dimen0=\wd0                                 
   \setbox1=\hbox{\sl/} \dimen1=\wd1            
   \ifdim\dimen0>\dimen1                        
      \rlap{\hbox to \dimen0{\hfil\sl/\hfil}}   
      #1                                        
   \else                                        
      \rlap{\hbox to \dimen1{\hfil$#1$\hfil}}   
      \hbox{\sl/}                               
   \fi}                                         %
%
\def\slashiii#1{\setbox0=\hbox{$#1$}#1\hskip-\wd0\hbox to\wd0{\hss\sl/\/\hss}}
%
\def\slashiv#1{#1\llap{\sl/}}

%
\def\dmu{\partial^{\mu}}
\def\dmul{\partial_{\mu}}
\def\dnu{\partial^{\nu}}
\def\dnul{\partial_{\nu}}
\def\dmudmu{\dmul\dmu}        \let\dalamb=\dmudmu
\def\Dmu{D^{\mu}}
\def\Dmul{D_{\mu}}
\def\contract#1#2{\gamma^{#1}{#2}_{#1}}
\def\del{\vector\nabla}
\def\delsq{\del^2}
\def\deldot#1{\del\mbf{\cdot}\vector #1}
%
%
%
%
%
%
\def\lagrangian{{\cal L}}        \let\lag=\lagrangian
\def\hamiltonian{{\cal H}}
\def\Amu{A^\mu}
\def\Amul{A_\mu}
\def\Fmunu{F^\mn}
\def\Fmunul{F_\mn}
\def\Gmunu{G^\mn}
\def\Gmunul{G_\mn}
\def\Vzero{V_{\sssize 0}}
\def\Vmu{V^{\mu}}
\def\Vmul{V_{\mu}}
\def\Vvec{\vector V}      \let\Vvector=\Vvec
\def\bzero{b_{\sssize 0}}
\def\bmu{b^{\mu}}
\def\bmul{b_{\mu}}
\def\Bvec{\vector B}
\def\psibar{\overline\psi}
\def\psidag{\psi^{\dagger}}         \let\psidagger=\psidag
\def\Gnk{G_{n\kappa}(r)}
\def\Fnk{F_{n\kappa}(r)}
\def\phizero{\phi_{\sssize 0}}
\def\Azero{A_{\sssize 0}}
\def\xzero{x_{\sssize 0}}
\def\ubar{\bar{u}}
\def\dbar{\bar{d}}
\def\cbar{\bar{c}}
\def\qbar{\bar{q}}
\def\qbarq{\qbar q}
\def\Tmunu{T^\mn}
\def\Tnumu{T^{\nu\mu}}
%
%
%
%
\def\intback{\kern-.1em}
\def\dthree#1{\intback{\rm d}^3\intback #1}
\def\dfour#1{\intback{\rm d}^4\intback #1}
\def\dthreepi#1{\intback {\tsize{\rm d}^3\intback #1 \over
                                 \mathstrut\tsize (2\pi)^3}}
\def\dfourpi#1{\intback {\tsize{\rm d}^4\intback #1 \over
                                 \mathstrut\tsize (2\pi)^4}}
\def\dn#1#2{\intback{\rm d}^{#1}\intback #2}
\def\dnpi#1#2{\intback {\tsize{\rm d}^{#1}\intback #2 \over
                                 \mathstrut\tsize (2\pi)^{#1}}}
\def\intdn#1#2{\int\dn#1#2}
\def\intdnpi#1#2{\int\dnpi#1#2}
\def\dfourq{\dfour{q}}
\def\dfourk{\dfour{k}}
\def\dfourx{\dfour{x}}
\def\dthreeq{\dthree{q}}
\def\dthreek{\dthree{k}}
\def\dthreex{\dthree{x}}
\def\intdfourkpi{\int\dfourpi{k}}
\def\intdfourqpi{\int\dfourpi{q}}
\def\intdnkpin{\int\intback {\tsize{\rm d}^n\intback k \over
                   \mathstrut\tsize (2\pi)^n}}
\def\dhpi#1{\intback {\tsize{\rm d}^4 \intback #1 \over
         \mathstrut\tsize (2\pi)^3}}
\def\partder#1#2{{\partial #1\over\partial #2}}
                    \let\partialderiv=\partder  \let\partderiv=\partder
\def\deriv#1#2{{{\rm d} #1\over {\rm d}#2}}
\def\dbyd#1{{{\rm d}\over {\rm d}#1}}
\def\partialdbyd#1{{\partial \over\partial #1}}   \let\partdbyd=\partialdbyd
%
%
%
%
\def\threej#1#2#3#4#5#6{\left(\matrix{#1 & #2 & #3 \cr
                                      #4 & #5 & #6 \cr}\right)}
\def\sixj#1#2#3#4#5#6{\left\{\matrix{#1 & #2 & #3 \cr
                                      #4 & #5 & #6 \cr}\right\}}
\def\ninej#1#2#3#4#5#6#7#8#9{\left\{\matrix{#1 & #2 & #3 \cr
                                           #4 & #5 & #6 \cr
                                           #7 & #8 & #9 \cr}\right\}}
\def\alphas{\alpha_{{\rm s}}}
\def\gv{g _{\rm v}}        \let\gvec=\gv
\def\gvsq{\gv^2}
\def\grho{g_\rho}
\def\grhosq{\grho^2}
\def\gs{g_{\rm s}}         \let\gsca=\gs
\def\gssq{\gs^2}
\def\gpi{g_{\pi}}
\def\gpisq{\gpi^2}
\def\fpi{f_{\pi}}
\def\fpisq{\fpi^2}
\def\Fpi{F_{\pi}}
\def\GF{G_{{\sssize \rm F}}}
\def\gpiNN{g_{\pi{\sssize NN}}}
%
%
%
%
\def\mq{m_{{\rm q}}}         \let\mquark=\mq
\def\mv{m _{\rm v}}      \let\mvec=\mv
\def\mvsq{\mv^2}
\def\mrho{m_\rho}
\def\mrhosq{\mrho^2}
\def\ms{m_{\rm s}}       \let\msca=\ms
\def\mssq{\ms^2}
\def\mpi{m_{\pi}}
\def\mpisq{\mpi^2}
\def\Mnucleon{M_{{\sssize N}}}  \let\mnucleon=\Mnucleon
%
%
%
%
%
\def\aspac{\noalign{\vskip 2pt}}
\def\bspac{\noalign{\vskip 4pt}}
\def\cspac{\noalign{\vskip 6pt}}
\def\dspac{\noalign{\vskip 8pt}}
\def\espac{\noalign{\vskip 10pt}}
\def\fspac{\noalign{\vskip 12pt}}
\def\gspac{\noalign{\vskip 14pt}}
\def\hspac{\noalign{\vskip 16pt}}
%
%
%
%
%
\def\nucleus#1#2{$^{#2}${\rm #1}}
%
%
%
%
%
\def\etc{{\it etc.}}
\def\etal{{\it et al.}}
\def\eg{{\it e.g.},\ }
\def\ie{{\it i.e.},\ }
\def\cf{{\it cf.}}
%
%
%
%
%
\def\Tav#1{\langle \mkern-3mu\langle \mkern2mu%
              #1 \mkern2mu \rangle \mkern-3mu\rangle}
     \let\ensavg=\Tav
\def\abs#1{| #1 |}
\def\overlap#1#2{\langle #1 | #2 \rangle}
\def\matrixelement#1#2#3{\bral{#1}#2\ketl{#3}}
   \let\me=\matrixelement
\def\commutator#1#2{[#1,#2]}
\def\anticommutator#1#2{\{#1,#2\}}
\def\sTav#1{\left\langle \mkern-3mu\left\langle \mkern2mu%
              #1 \mkern2mu \right\rangle \mkern-3mu\right\rangle}
\def\sabs#1{\left| #1 \right|}

\def\stdket#1{\left. #1 \right\rangle}
\def\soverlap#1#2{\bra{#1\vphantom{#1#2}}\stdket{#2\vphantom{#1#2}}}

\def\smatrixelement#1#2#3{\bra{#1\vphantom{#2#3}}#2\ket{#3\vphantom{#1#2}}}
   \let\sme=\smatrixelement
\def\scommutator#1#2{\left[#1,#2\right]}
\def\santicommutator#1#2{\left\{#1,#2\right\}}
\def\VEV#1{\left\langle #1 \right\rangle}    \let\vev=\VEV
\def\VEVl#1{\langle #1 \rangle}
\def\VEVgl#1{\bigl\langle #1 \bigr\rangle}
\def\VEVGl#1{\Bigl\langle #1 \Bigr\rangle}
\def\VEVggl#1{\biggl\langle #1 \biggr\rangle}
\def\bra#1{\left\langle #1 \right|}
\def\ket#1{\left| #1 \right\rangle}
\def\bral#1{\langle #1 |}
\def\ketl#1{| #1 \rangle}
\def\bragl#1{\bigl\langle #1 \bigr|}
\def\ketgl#1{\bigl| #1 \bigr\rangle}
\def\Bragl#1{\Bigl\langle #1 \Bigr|}
\def\Ketgl#1{\Bigl| #1 \Bigr\rangle}
\def\braggl#1{\biggl\langle #1 \biggr|}
\def\ketggl#1{\biggl| #1 \biggr\rangle}
\def\dubdots{}
\def\dubbral#1{\langle #1 \Vert}
\def\ddubbral#1{\langle #1 \dubdots}
\def\dubketl#1{\Vert #1 \rangle}
\def\ddubketl#1{\dubdots #1 \rangle}
\def\dubbragl#1{\bigl\langle #1 \bigr\Vert}
\def\ddubbragl#1{\bigl\langle #1 \dubdotsgl}
\def\dubketgl#1{\bigl\Vert #1 \bigr\rangle}
\def\ddubketgl#1{\dubdotsgl #1 \bigr\rangle}
\def\dubBragl#1{\Bigl\langle #1 \Bigr\Vert}
\def\ddubBragl#1{\Bigl\langle #1 \dubDotsgl}
\def\dubKetgl#1{\Bigl\Vert #1 \Bigr\rangle}
\def\ddubKetgl#1{\dubDotsgl #1 \Bigr\rangle}
\def\dubbraggl#1{\biggl\langle #1 \biggr\Vert}
\def\ddubbraggl#1{\biggl\langle #1 \dubdotsggl}
\def\dubketggl#1{\biggl\Vert #1 \biggr\rangle}
\def\ddubketggl#1{\dubdotsggl #1 \biggr\rangle}
\def\dubbra#1{\left\langle #1 \left\Vert}
\def\dubket#1{\right\Vert #1 \right\rangle}
%
%

%
%
%
%
\def\Phiket#1{\ket{\Phi_{#1}}}
\def\Phibra#1{\bra{\Phi_{#1}}}
\def\Psiket#1{\ket{\Psi_{#1}}}
\def\Psibra#1{\bra{\Psi_{#1}}}

\newcommand{\finalnewpage}{\newpage}
\def\Cv{C _{\rm v}}        
\def\Cvsq{\Cv^2}
\def\Crho{C_\rho}
\def\Crhosq{\Crho^2}
\def\Cs{C_{\rm s}}         
\def\Cssq{\Cs^2}
\def\kFermi#1{k_{\sssize {\rm F}}^{#1}}    
\def\rhoBto#1{\rho_{\sssize {\rm B}}^{#1}} 

\title{Building Atomic Nuclei with the\\
       Dirac Equation}\thanks{%
Presented at the Dirac Centennial Symposium, Florida State University,
December 6--7, 2002.}

\author{Brian D. Serot}\email{serot@iucf.indiana.edu}
\affiliation{Department of Physics and Nuclear Theory Center
             Indiana University, Bloomington, IN\ \ 47405}
\author{\null}
\noaffiliation

\date{March, 2003}

\begin{abstract}
The relevance of the Dirac equation for computations of nuclear structure
is motivated and discussed.
Quantitatively successful results for medium- and heavy-mass nuclei are
described, and modern ideas of effective field theory and density functional
theory are used to justify them.
\end{abstract}

\maketitle

\section{Introduction}
\label{sec:Intro}

To understand how to build atomic nuclei with the Dirac equation,
we will begin by asking some simple questions.
What are the basic nuclear properties that we are trying to 
correlate and predict?
Why use hadrons (rather than quarks and gluons) as the degrees of freedom?
Why use the Dirac equation rather than the Schr\"odinger equation
to describe the dynamics?
How can we build a simple model of nuclear matter that reproduces
the empirical equilibrium properties and that can be extended
to calculations of medium- and heavy-mass nuclei?
How does the Dirac approach predict the nuclear shell model?
And how can we relate the hadronic description of nuclei to the
underlying strong-interaction theory of quantum chromodynamics
(QCD)?

The basic properties of nuclei provide stringent constraints
on any nuclear theory.
An accurate description of these properties is necessary for
any useful predictions or extrapolations.
We will concern ourselves primarily with bulk and 
single-particle nuclear properties, as listed below; a more
detailed discussion can be found in 
Refs.~\cite{FW,subatomic}.

We certainly want to reproduce the observed shapes of nuclei:
the interior density of a heavy nucleus should be relatively
constant, since the nuclear forces ``saturate'' at
the equilibrium density of nuclear matter
$(\textrm{roughly}\ \rhozero \approx 0.15\, {\rm fm}^{-3})$.
Moreover, the nucleus should have a well-defined surface, with
the density decreasing from 90\% to 10\% of its central value
over a distance of roughly 2 fm.
Finally, because of saturation,
the radius $R$ of a nucleus should scale according to
$R \approx A^{1/3}\, 1.1\, {\rm fm}$, where $A = N + Z$ is the
total number of neutrons ($N$) plus protons ($Z$).

The total energy of the nucleus should follow the ``liquid drop''
formula
\begin{equation}
E = -{a_1}A + {a_2}A^{2/3}+  {a_3}Z^2/A^{1/3}
                  + {a_4}(N - Z)^2/A + \cdots \ ,
\label{eq:liquiddrop}
\end{equation}
where typical values for the $a_i$ coefficients are given in
Ref.~\cite{FW}.

The particle spectrum is determined by the qualitative
features of the single-particle potential.
In a nonrelativistic (Schr\"odinger) language, the central 
potential is midway between a harmonic oscillator and a square
well; this shape determines the ordering of the levels as 
a function of the orbital angular momentum.
(See Ref.~\cite{FW}, Figs.~57.1 and 57.2.)
In addition, the spin-orbit potential is strong, which is
instrumental in determining the major shell closures and, hence,
the shell model.
We will see below how these features are reproduced in a description
based on the Dirac equation.

These simple nuclear features are the ones we will focus on.
We expect that they can be described adequately by a single-particle
equation with an effective, one-body interaction.
Such an approach has many names, depending on the system being
studied and on the practitioner: ``shell model'', ``mean-field
theory'', ``Kohn--Sham'' density functional theory, etc.
Our goal is to correlate (fit) a modest number of nuclear bulk and 
single-particle data and then to predict other, similar data as
well as possible.

\section{Why Use Hadrons?}
\label{sec:hadrons}

Well, why not?
Our focus is on long-range nuclear characteristics, and all measured
observables are colorless.
(In fact, most of the observables relevant to us are dominated by
the \textit{isoscalar} part of the interaction.)
Moreover, hadronic variables (baryons and mesons) are efficient, since
hadrons are the particles that are observed in experiments.
Colored quarks and gluons participate 
\textit{only in intermediate states},
and such ``off-shell behavior'' is unobservable; by using hadrons, we
expend no theoretical effort combining quarks and gluons into color
singlets that can actually be observed.

So we pick the most efficient degrees of freedom by choosing hadrons.
We will have to parametrize the nuclear hamiltonian anyway, since
we cannot compute its true form from QCD, and hadronic variables, 
if combined in all forms consistent with the underlying
symmetries, provide sufficient flexibility for our parametrization.
We cannot guarantee that a single-particle hadronic approach will be
successful in describing the observables of interest, but we want to
see how well we can do.

\section{Why Use the Dirac Equation?}
\label{sec:Dirac}

To motivate the Dirac equation as straightforwardly as possible, 
compare the particle spectrum (and fine structure) in a light atom 
with the spectrum in a heavy nucleus.
An example of the former is given in Ref.~\cite{BjD}, 
while an early example of
the latter is given in Ref.~\cite{Mayer55}, 
which is reproduced in Fig.~57.3 of Ref.~\cite{FW}.
The most striking result is that it is impossible to draw the atomic
fine structure to scale, since the splittings are roughly 1/10,000
as large as the major-level splittings (at least for the deeply bound
atomic levels).
In contrast, the nuclear spectrum shows that the ``fine'' structure
is really ``gross''; the fine-structure splittings are as large as
the major-level splittings to within a factor of two!

The implication is that there \textit{must\/}
be some relativistic effects that
are \textit{important\/} in nuclei (unlike light atoms), and thus
it is much more natural to use the Dirac equation to describe the
quasi-particle nucleon wave functions.
We will now try to understand this result by building a simple
model of uniform nuclear matter.

\section{A Simple Model of Nuclear Matter}
\label{sec:Walecka}

We consider a model first proposed by Walecka \cite{Walecka74},
which contains nucleons $(\psi )$ and neutral (isoscalar) Lorentz
scalar $(\phi )$ and vector $(V^{\mu} )$ mesons.
This model is often referred to as ``quantum hadrodynamics I''
(or QHD--I, for short).
The lagrangian density for this model (using the conventions of
Ref.~\cite{IJMPE} and suppressing counterterms for simplicity) is
\begin{eqnarray}
\mathcal{L} & = & \psibar (i \gamma_{\mu} \partial^{\mu} - M) \psi
          + \frac{1}{2} ( \partial_{\mu} \phi \, \partial^{\mu} \phi
              - \mssq \phi^{2} ) 
             - \frac{1}{4} (\partial_{\mu} V_{\nu} - \partial_{\nu}
             V_{\mu})^2  \nonumber\\[3pt]
   & & \quad  {} + \frac{1}{2} \mvsq V_{\mu} V^{\mu}
       + \gs \psibar \psi \phi - \gv V^{\mu} \, \psibar\gammamul\psi
     \ .
\label{eq:QHDI}
\end{eqnarray}
The included degrees of freedom are the minimal ones that will allow
us to understand the qualitative features of the nuclear many-body
system, which is our goal.
We will describe the system in terms of Dirac quasi-particles moving
in classical meson mean fields, an approximation that we will elaborate
on and justify later.
Note that the baryon current (density) $\psibar\gammamul\psi$ is conserved.

It is important to emphasize that the Lorentz scalar and vector fields
are \textit{effective fields} that are introduced to parametrize the
nucleon--nucleon (NN) interaction.
The quanta of these fields never appear ``on the mass shell'' as real 
particles in any of the calculations discussed here.
They are analogous to the phonons that describe electron--electron
interactions inside a metal.

If one computes the NN interaction using one-boson exchange [purely for
illustration, since the coupling constants $\gs$ and $\gv$
in Eq.~(\ref{eq:QHDI}) are
large], one finds a short-range repulsion (from $V^{\mu}$) and a mid-range
attraction (from $\phi$), which is characteristic of the NN force.
Explicit pion exchange is of minor importance for our observables of 
interest; isoscalar, scalar and vector fields are dominant for the bulk
and single-particle properties of heavy nuclei (and some 
\textit{multi-pion\/} exchange is simulated by our effective fields anyway).

Consider nuclear or neutron matter at zero temperature.
We can treat the mesons at the mean-field level by taking
\begin{equation}
\phi \longrightarrow \langle \phi \rangle \equiv \phi_0 \ , \quad
\Vmu \longrightarrow  \langle \Vmu \rangle \equiv V_0 \, \delta^{\mu 0} \ ,
\end{equation}
where $\langle \textbf{V} \rangle \equiv 0$, 
since we assume that we are in the rest frame
of the uniform matter.
Note that $\phi_0$ and $V_0$ are \textit{constants}.

Why should mean meson fields yield a reasonable description of the system,
since QHD--I is a strong-coupling theory?
Our goal is to construct an approximate energy functional of the scalar
$(\rhos )$ and baryon $(\rhoB )$ densities and to fit the parameters in
this functional to bulk nuclear properties.
The mean meson fields give us a convenient way to do this, since they
satisfy the \textit{mean-field equations}:
\begin{equation}
 \phi_0 = \frac{\gs}{\mssq} \, \rhos 
         \equiv \frac{\gs}{\mssq} \sum_i^{\rm occ}
         \, {\psibar}_i {\psi}_i \ , 
\label{eq:phimft}
\end{equation}
\begin{equation}
  V_0 = \frac{\gv}{\mvsq} \, \rhoB 
        \equiv \frac{\gv}{\mvsq} \sum_i^{\rm occ}
         \, \psi^{\dagger}_i {\psi}_i \ , 
\label{eq:Vmft}
\end{equation}
\begin{equation}
  [\, i \gammamul \partial^{\mu} - \gv \gamma^0 V^0 -
         \underbrace{( M - \gs \phi_0)}_{\Mstar} \, ] \, \psi
         = 0 \ ,
\end{equation}
where ``occ'' signifies the occupied quasi-particle levels.
(In infinite matter, we sum over states with both spin projections
and with momentum
$k \le \kfermi$, where $\kfermi$ is the Fermi momentum.)

We solve these equations for stationary quasi-particle states; the problem
is self-consistent, since $\rhos = \rhos (\Mstar )$ both determines and
depends on the wave functions \cite{Walecka74,IJMPE}.
\ The nuclear/neutron matter energy function(al) then becomes
\begin{equation}
\mathcal{E}_{\rm MFT} =  \frac{\gvsq}{2 \mvsq}\, \rhoBto{2}
                + \frac{\mssq}{2 \gssq} \, (M - \Mstar )^2 
   {} + \frac{\lambda}{\pi^2} 
    \int_0^{\kfermi} \! {\rm d}t \; t^2 \, (t^2 + \Mstarsq )^{1/2}
              \ ,  \label{eq:edens}
\end{equation}
where the baryon density is
\begin{equation}
\rhoB \equiv \frac{\lambda \kFermi{3}}{3 \pi^2} \ , 
\end{equation}
and the isospin degeneracy is $\lambda = 2$ for 
symmetric $(N = Z)$ nuclear matter
and $\lambda = 1$ for pure neutron matter $(Z=0)$.
(Note that, by definition, the Coulomb force between protons
is turned \textit{off}.)

One can now minimize the energy density $\mathcal{E}$ with respect to
$\rhoB$ to find the equilibrium point, and use the \textit{empirical\/}
equilibrium point of nuclear matter
(density = 
$\rhozero \approx 0.15\,\textrm{fm}^{-3}$, binding energy =
$e_{\sssize 0} \approx 16\,\textrm{MeV})$ to determine the two unknown
ratios
%
$$   \frac{\gssq}{\mssq} \quad {\rm and} \quad
   \frac{\gvsq}{\mvsq} \ , $$
%
which are expressed more conventionally (and less dimensionally) as
\begin{equation}
\Cssq \equiv \frac{\gssq M^2}{\mssq} = 357.4 
\quad {\rm and} \quad
\Cvsq \equiv \frac{\gvsq M^2}{\mvsq} = 273.8 \ .
\end{equation}

The resulting nuclear/neutron matter binding curves and the self-consistent
effective mass $\Mstar$ as functions of the density are shown in Figs.~1
and 2 of Ref.~\cite{IJMPE}.
The important features of these results are:
\begin{itemize}
\item
Symmetric nuclear matter is a self-bound liquid with an equilibrium point
as defined above.
This illustrates the ``saturation'' of nuclear forces.
\item
Pure neutron matter is (generally) unbound at all densities.
This reflects the positive symmetry-energy coefficient [$a_4$ in 
Eq.~(\ref{eq:liquiddrop})] that enters when the number of neutrons
and protons is different.\footnote{%
In QHD--I, this coefficient is too small.
One must add a $\rho$ meson, which couples to the difference of
proton and neutron densities, to achieve an accurate result.
See Ref.~\protect\cite{IJMPE}.}
\item
The nucleon effective mass at equilibrium density is roughly 
$\Mstar_0 \approx 0.6 M$.
This shows that the scalar mean field is roughly $-400$ MeV at equilibrium;
the corresponding vector mean field is roughly 300 MeV, and the two fields
cancel to produce the relatively small binding energy of 16 MeV.
We turn now to a discussion of this point.
\end{itemize}

What causes the nuclear matter saturation and the relatively small binding
energy?
Let's expand $\mathcal{E}/ \rhoB$ from Eq.~(\ref{eq:edens}) in powers of
$\kfermi$ \cite{IJMPE}:
%
\begin{eqnarray}
\mathcal{E}_{\rm MFT} / \rhoB & = & M + 
    \left[ \frac{3 \kFermi{2}}{10 M}
          -\frac{3 \kFermi{4}}{56 M^3}
          +\frac{  \kFermi{6}}{48 M^5} - \cdots \right] 
          + \frac{\gvsq}{2 \mvsq} \, \rhoB 
          - \frac{\gssq}{2 \mssq} \, \rhoB \nonumber\\[3pt]
& &  \quad
     {} + \frac{\gssq}{\mssq} \, \frac{\rhoB}{M}
       \left[ \frac{3  \kFermi{2}}{10 M}
        -\frac{36 \kFermi{4}}{175 M^3} + \cdots \right] 
     + \left( \frac{\gssq \rhoB}{\mssq M} \right)^2
       \left[ \frac{3 \kFermi{2}}{10 M} - \cdots \right]
     \nonumber\\[3pt]
& &  \quad
     + \left( \frac{\gssq \rhoB}{\mssq M} \right)^3
       \left[ \frac{3 \kFermi{2}}{10 M} - \cdots \right] 
     + \cdots 
\label{eq:expansion}
\end{eqnarray}
The lowest-order Lorentz scalar and vector contributions (which are
proportional to $\rhoB$) \textit{set the scale\/} for the large
mean fields.
[See Eqs.~(\ref{eq:phimft}) and (\ref{eq:Vmft}).]
This scale is consistent with chiral QCD counting rules \cite{VMD,Params},
but these two terms \textit{cancel almost exactly\/} in the binding energy,
leading to an anomalously small remainder.
However, they \textit{add constructively\/} in the spin-orbit interaction,
leading to appropriately large
spin-orbit splittings in nuclei \cite{Furry36,HS81,spinorbit}.

It is important to notice the \textit{different behavior\/} of the vector
and scalar interaction terms in Eq.~(\ref{eq:expansion}).
Whereas the vector interaction enters at only linear order in $\rhoB$,
the scalar interaction enters at all orders; moreover, the leading scalar
term at every order in $\rhoB$ looks exactly the same, and they all add
constructively.
These terms are precisely what one gets by shifting the nucleon mass in
the nonrelativistic kinetic energy term $3 \kFermi{2}/10 M$ 
from $M \rightarrow \Mstar \approx M - \gssq \rhoB / \mssq$.
These additional, repulsive, velocity-dependent interactions reduce the
strength of the lowest-order, attractive scalar contribution and are crucial
for establishing the location of the equilibrium point of nuclear matter.
Thus the different behavior of the vector and scalar interactions leads
to \textit{large relativistic interaction effects\/}
in the nuclear matter energy density.
In contrast, the relativistic corrections to the kinetic energy (the 
nonleading terms in the first pair of square brackets) are indeed small; 
this is \textit{not\/} where the important ``relativity'' is.

\section{Mean-Field Theory for Nuclei}
\label{sec:nuclei}

We now compute the bulk and single-particle
properties of atomic nuclei using essentially the same simple lagrangian
discussed above.
Our treatment follows that of Ref.~\cite{HS81}, which is more than
twenty years old, but which is still sufficient to illustrate the important
points.
We will discuss modifications and more modern treatments later.

The basic idea is to allow the mean meson fields to be spatially
dependent, and we will consider only spherically symmetric nuclei for
simplicity.
We again look for stationary quasi-nucleon states, and so the mean-field
equations become \cite{IJMPE}
\begin{equation}
\del^2 \phi_0 (r) - \mssq \phi_0 (r) = - \gs \rhos (r) \ , 
\end{equation}
\begin{equation}
\del^2 V_0 (r) - \mvsq V_0 (r) = - \gv \rhoB (r) \ , 
\end{equation}
\begin{equation}
\{ {-i}\, \bm{\alpha} \bm{\cdot} \del + \gv V_0 (r)
   + \beta [M - \gs \phi_0 (r)] \} \,
   \psi_{\alpha} (\textbf{x}) = \epsilon_{\alpha} \psi_{\alpha} 
   (\textbf{x}) \ .
\end{equation}
These are coupled, nonlinear, differential equations that must be solved
\textit{self-consistently}.
They are sometimes called Dirac--Hartree equations \cite{Serot86,IJMPE}
but are more accurately described as Kohn--Sham equations \cite{Comments},
as we discuss in more detail below.

As one might expect, an accurate description of nuclear properties is not
possible using only nucleons and isoscalar mesons.
One must extend the model to include at least the Coulomb interaction 
between protons and an isovector $\rho$ meson that allows for a more
realistic description of the nuclear symmetry energy.
(See Refs.~\cite{HS81,Serot86} for details.)
The augmented model now contains four adjustable parameters:
$$ \gs \ , \quad \gv \ , \quad \grho \ , \quad \ms \ .$$
(The heavy meson masses are fixed at some ``large'' mass scale that is
roughly equal to the nucleon mass $M$.)
The couplings are fitted to the equilibrium point of symmetric nuclear
matter and to the nuclear matter symmetry energy; the length scale, which
is determined by $\ms$, is set by fixing this parameter to reproduce
the rms charge radius of a doubly magic nucleus, such as \nucleus{Ca}{40}.

Many nuclear structure calculations have been carried out within this
relativistic mean-field theory (RMFT) framework.
(See, for example, Refs.~\cite{HS81,Serot86}] or the 
extensive list of references in Ref.~\cite{IJMPE}.)
One finds that the bulk properties of nuclei are well reproduced even
in this relatively simple mean-field theory.
Moreover, the single-particle spectrum reveals the well-known nuclear
shell structure; this comes \textit{for free}, since the parameters are
fitted to the \textit{bulk\/} properties of nuclear matter (and one 
nuclear length scale).

Extensions of this simple model have been made to ``fine-tune'' the
results.  In the 1980's, numerous authors added terms involving nonlinear
interactions of the scalar field:
\begin{equation}
\mathcal{L}' = {} -\frac{1}{3!} \, \kappa \, \phi^3
               -\frac{1}{4!} \, \lambda \, \phi^4 \ ,
\end{equation}
and in the 1990's, various practitioners added vector self-couplings, like
\begin{equation} 
\mathcal{L}'' = {} + \frac{1}{4!} \, \zeta \, \gv^4 (V_{\mu} V^{\mu} )^2 \ ,
\end{equation}
as well as other nonlinear and gradient-coupling terms, some motivated
by the ideas of effective field theory; see the discussion below.
(Many calculations in these extended models are cited in
Ref.~\cite{IJMPE}.
For an alternative approach that uses only nucleons in a lagrangian that
contains numerous powers of fermion fields, see Ref.~\cite{Rusnak97}
and references therein.)

These additional nonlinearities can be interpreted in terms of many-body
nuclear forces, and they introduce additional density dependence into the
nuclear energy functional, which allows it to more accurately reproduce
the true energy functional.
The new parameters are fitted either to additional nuclear matter 
properties, or to other theoretical calculations of nuclear matter (based
on the Schr\"odinger equation), or to a selected set of data from finite
nuclei.
The basic conclusion from these extended calculations is that the successful
qualitative features predicted by the original simple models persist,
but the quantitative accuracy increased by nearly two orders of magnitude 
over a period of twenty years.
For a comparison of the accuracy of results obtained with different
collections of parameters, see, for example, 
Refs.~\cite{VMD,Params}.
For some recent state-of-the-art \textit{predictions\/} of this approach
(that is, calculations of nuclei that are not included in the
fitting procedures), see Ref.~\cite{Huertas02}.

But it still remains for us to understand at a deeper level why these
simple relativistic mean-field calculations can do such an excellent job
of reproducing certain nuclear observables.
For this, we must study \dots

\section{Modern Developments}
\label{sec:modern}

The discussion in this section is a synopsis of the formalism presented in 
Refs.~\cite{IJMPE,VMD,Comments,Kohn99}, 
which is based on the ideas of modern effective field theory (EFT) and 
density functional theory (DFT).
The interpretation of the earlier, successful results using EFT/DFT puts
them on a firm theoretical basis.

First of all, we interpret QHD as a nonrenormalizable EFT.
This means that it contains \textit{known long-range interactions\/} that
are constrained by the underlying QCD symmetries, plus a complete (but
non-redundant) set of \textit{generic short-range interactions}, i.e.,
``contact'' and ``gradient'' terms.
The borderline between short and long ranges is characterized by the
\textit{breakdown scale\/} $\Lambda$
of the EFT; empirically, we find that $\Lambda \approx 600\,\textrm{MeV}$
for QHD \cite{Params}.

If we ignore strangeness, then only nucleons and pions are ``real''
(stable) particles.
The other field quanta are always virtual 
and just let us parametrize the NN interaction.
As in any lagrangian theory, there are different ways to choose the
generalized coordinates (fields), but some coordinates may be more efficient 
than others \cite{Params,Rusnak97}.

The QHD EFT lagrangian explicitly exhibits the symmetries of QCD:
The global, chiral $\textrm{SU}(2)_L \times \textrm{SU}(2)_R$ 
symmetry is nonlinear,
approximate, and spontaneously broken \cite{IJMPE}.
\ The remaining global, isovector subgroup $\textrm{SU}(2)_V$ is realized
linearly.
It is straightforward (but usually tedious) to include electromagnetic
interactions through the familiar local U(1) gauge symmetry \cite{VMD}.

The basic strategy for using the QHD lagrangian has been developed over
the last several years \cite{Bodmer,VMD,IJMPE}.
\ First, assign an index $\nu$ to each term in the lagrangian:
\begin{equation}
\nu = d + n/2 + b \ ,
\end{equation}
where $d$ is the number of derivatives (not counting those that act on
nucleon fields),\footnote{Time derivatives 
acting on nucleon fields will generally bring down factors of the
nucleon mass or energy, which are not small compared to $\Lambda$.}
$n$ is the number of nucleon fields, and $b$ is the number of
non-Goldstone bosons.

Now organize the lagrangian in powers of $\nu$ and truncate.
This gives an expansion in inverse powers of a heavy mass scale
$\Lambda \approx M$, which has been shown to be reliable in calculations
of medium- and heavy-mass nuclei \cite{VMD,Params}.
Practically speaking, in the nuclear many-body problem, this expansion is
in powers of $\kfermi /M$, where $\kfermi$ is the Fermi momentum at
equilibrium nuclear density $(\kfermi /M \approx 1/3)$.

Use the truncated lagrangian to construct an energy functional, which
is to be interpreted within the DFT framework:
We approximate the functional using \textit{factorized\/} densities or
fields, which produces a mean-field form of the functional.
Expand it as a power series in density and momentum (by counting powers of
$\nu$) and fit the remaining parameters to a restricted set of experimental
data \cite{Params}.
These data typically include nuclear binding energies, prominent features
of the nuclear electromagnetic charge form factors, and single-particle
energy splittings for the least-bound orbitals \cite{VMD}.
Define a set of Kohn--Sham (KS) single-particle orbitals that satisfy
differential equations obtained by extremizing the energy functional with
respect to the densities and fields.
This procedure guarantees that all of the source terms in these equations
are \textit{local}.
The KS orbitals are tailored to the generation of the ground-state density,
and they include short-range and correlation effects adequately, if the
mean-field energy functional is a good approximation to the true energy
functional \cite{Kohn99}.

The mean-field energy functional constructed above omits some
long-range contributions, which are generally nonlocal and nonanalytic
functions of the densities.
These contributions can be added systematically, by computing loop integrals
using the well-known rules of EFT \cite{Hu00}.
The effect of these loop contributions on the energy functional of
atomic nuclei is an important topic for future study.

\section{Summary}
The most important points in the preceding discussion can be summarized
as follows:
\begin{itemize}
\item
The Dirac equation provides an economical and natural way to describe bulk
nuclear properties and the nucleon single-particle spectrum, 
with the correct
spin-orbit force (that is, the nuclear shell model) arising 
\textit{automatically}.
\item
Kinematical relativistic effects are small in nuclei, but dynamical
relativistic effects from the interactions are important.
\item
Modern QHD EFT's incorporate the basic symmetries of QCD.
\item
The mean-field approach to heavy nuclei is really DFT, implemented through
KS quasi-particle orbitals.
The tested validity and accuracy of our truncation procedure for both 
fitted \textit{and predicted\/} results shows that we really know something
about the energy functional for cold nuclear matter near equilibrium density!
\item
The energy functional can be extended beyond the mean-field parametrization
using well-defined rules of EFT to compute the long-range contributions
of loop integrals.
This has been done recently \cite{Hu00}.
\item
The QHD/EFT/DFT/KS formalism provides a \textit{true representation\/}
of QCD in the low-energy nuclear domain.
\end{itemize}

The basic message of this talk is: the Dirac equation is relevant for 
nuclear-structure physics, even though you might not expect it to be.
Our quantitative successes justify its usage, but the modern theoretical
ideas of EFT and DFT explain why it works.

\section*{Acknowledgments}
I am grateful to the organizers of the Dirac Centennial Symposium for the
opportunity to visit Tallahassee and to share my ideas about the structure
of atomic nuclei.
I am also pleased to acknowledge long, friendly, and fruitful collaborations
with Dick Furnstahl, Ying Hu,
Hua-Bin Tang, and Dirk Walecka during the course
of these studies.
This work was supported in part by the US Department of Energy under
contract no.~DE--FG02--87ER40365.

\finalnewpage

\end{document}